# Torrefaction of Pelletized Oil Palm Empty Fruit Bunches

Bemgba Bevan Nyakuma[a*], Arshad Ahmad[a], Anwar Johari[a], Tuan Amran Tuan Abdullah[a], Olagoke Oladokun[a]

*a:Institute of Hydrogen Economy, Faculty of Chemical Engineering,
Universiti Teknologi Malaysia, 81310 UTM Skudai, Johor Bahru, Malaysia.
Corresponding email: bnbevan2@live.utm.my*

**Abstract**

The torrefaction of oil palm Empty Fruit Bunch (EFB) briquettes was examined in this study. The results revealed that temperature significantly influenced the mass yield, energy yield and heating value of EFB briquettes during torrefaction. The solid uniform compact nature of EFB briquettes ensured a slow rate of pyrolysis or devolatization which enhanced torrefaction. The mass yield decreased from 79.70 % to 43.03 %, energy yield from 89.44 % to 64.27 % during torrefaction from 250 °C to 300 °C. The heating value (HHV) of oil palm EFB briquettes improved significantly from 17.57 MJ/kg to 26.24 MJ/kg after torrefaction at 300 ºC for 1 hour. Fundamentally, the study has highlighted the effects of pelletization and torrefaction on solid fuel properties of oil palm EFB briquettes and its potential as a solid fuel for future thermal applications.

1.  **Introduction**

The emission of anthropogenic greenhouse gases has increased at an average rate of 2.1 % over the years particularly due to fossil fuel combustion, rapid deforestation, and environmental pollution. This scenario has increased the urgent need for a sustainable transition from fossil fuels to renewables. Biomass is promising renewable energy technology (RET) with the potential to convert waste streams into chemicals, fuels, and power for the future [1-3]. Currently, the production of energy from solid fuels is dominated by coal, which has superior fuel properties compared to biomass [4].

With over 140 million tons of oil palm waste generated annually, Malaysia can potentially generate 24 % of its current primary energy supply. Hence, the availability of large quantities of lignocellulosic oil palm waste in Malaysia presents a unique opportunity for the future generation of clean renewable energy [5]. Biomass utilization for energy production is hindered by high moisture content, alkali content, bulky heterogeneous nature, low heating value and conversion efficiencies [5, 6]. The outlined properties of biomass present technical challenges particularly in the design of conversion equipment and efficient utilization of solid fuels [4].

Pre-treatment techniques such as pelletization and torrefaction can improve the utilization and conversion efficiency of biomass resources [7]. Pelletization is a densification technique used to improve the thermochemical conversion efficiency, handling, and transport of biomass resources [8]. Torrefaction is a mild pyrolysis process in which biomass is treated in the temperature range from 200 °C - 300 °C in a non-oxidising environment. The drying and partial devolatization of biomass during torrefaction decreases the mass of biomass without affecting the energy content [9]. Furthermore, torrefaction improves the solid fuel properties of biomass by depolymerising long chain polysaccharide and removing $CO_2$ and $H_2O$. Hence by reducing the oxygen-to-carbon (O/C) ratio, the energy density and hygroscopic nature of biomass is significantly enhanced [6, 9]. Furthermore, the modified structure of torrefied biomass improves its grindability, gasification potential and future prospects as a solid fuel for co-firing in existing coal power plants [10-12]. Principally, the fuel properties of torrefied biomass is significantly influenced by torrefaction temperature, time, and type of biomass [9]. A number of studies have explored the torrefaction of agricultural waste [5, 7], wood [13-15], and energy grasses [4, 16]. Uemura and co-workers [5, 7] examined the effects of torrefaction temperature and oxygen concentration on the mass and energy yields of oil palm waste from 220 °C - 300 °C. The findings indicated that the torrefaction of oil palm Empty Fruit Bunches (EFB) at 300 °C for 30 mins resulted in an average mass yield of 24 % and energy yield of 56 %. In addition, the heating value (HHV) of torrefied EFB was 20.41 MJ/kg, which is lower than lignite coal currently utilised for energy production in coal power plants. Therefore, the poor fuel properties and low efficiencies of current oil palm waste conversion techniques require further investigation to improve the product yield of pre-treatment and its overall prospects as an ideal solid fuel for future thermal energy applications in Malaysia.

Hence, this study is aimed at investigating the torrefaction of oil palm empty fruit bunch (EFB) briquettes as a potential fuel for future thermal applications. The effect of torrefaction temperature on the mass yield, energy yield, energy density and heating value on EFB briquettes will be presented in detail using a simple and practical laboratory technique not previously reported in literature.

## 2. Experimental

The EFB briquettes used in this study were obtained from Felda Semenchu Sdn Bhd, Johor, Malaysia and used without treatment. For each experimental run, 10 g of the EFB briquettes were tightly enclosed in aluminium foil to prevent contact with air or oxygen, placed in a covered ceramic crucible and torrefied at 250 °C, 275 °C, and 300 °C for 1 hour in a muffle furnace (Model, Ney Vulcan D-130) as presented in Figure 1.

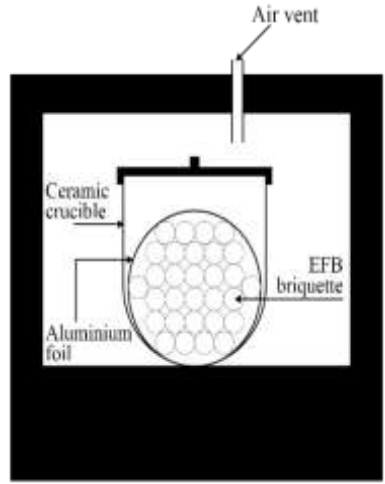

**Fig 1**. Experimental Set Up.

The experiments were repeated three times to establish the reliability and repeatability of the results. The simplicity and practicality of the proposed torrefaction technique is emphasized by the fact that it eliminates the need for nitrogen gas and other non-oxidising ancillary equipment reported in literature. Furthermore, the emphasis in literature is on biomass torrefaction followed by pelletization; whereas our technique proposes pelletization and subsequently torrefaction. The proximate analysis of the EFB briquette was determined using ASTM standard techniques with the following results; moisture content, 8.04 %; volatiles, 72.37 %; Fixed carbon 14.38 %; and Ash, 5.21 %. The elemental analysis of the EFB Briquettes and torrefied products were analysed using the vario MICRO cube CHNS/O elemental analyser. The XRF analysis was carried out using the Bruker S4 Pioneer X-Ray Fluorescence Spectrometer. The heating value (HHV) of the EFB briquettes and torrefied products was measured using a bomb calorimeter (Model, IKA C2000). The measurements were repeated three times in each case to ensure repeatability. The mass and energy yield of torrefaction are defined by the relations [4];

$$y_{mass} = \left(\frac{mass\ after\ torrefaction}{mass\ of\ sample\ before\ torrefaction}\right) \times 100 \qquad (1)$$

$$y_{energy} = y_{mass} \times \left(\frac{heating\ value\ (HHV)\ of\ torrefied\ product}{heating\ value\ (HHV)\ of\ sample\ before\ torrefaction}\right) \times 100 \qquad (2)$$

$$y_{energy\ desity} = \left(\frac{y_{energy}}{y_{mass}}\right) \qquad (3)$$

## 3. Results And Discussion

**Yield of Torrefaction**

Table 1 presents the mass yield, energy yield and energy density of the torrefied EFB briquettes. As observed, mass yield decreased from 79.70 % to 43.03 %, while the energy yield decreased from 89.44 % to 64.27 % during torrefaction. The decrease in mass and energy yield during torrefaction is primarily due to drying, partial devolatization and the breakdown of hemicellulose [9]. In addition, the results showed that temperature significantly influenced the mass and energy yield of oil palm EFB briquettes during torrefaction, a trend also reported by [4, 6, 7].

**Table 1.** Product yield of Oil palm EFB briquettes torrefaction.

| Temperature (ºC) | Mass Yield (%) | HHV (MJ/kg) | Energy yield (%) | Energy Density |
|---|---|---|---|---|
| 250 | 79.70 | 19.72 | 89.44 | 1.12 |
| 275 | 61.40 | 21.35 | 74.63 | 1.22 |
| 300 | 43.03 | 26.24 | 64.27 | 1.49 |

To analyse the effect of pelletization on the product yield of torrefaction, we compared our results with the mass yield of EFB obtained by *Uemura et al., 2011* by also carrying out torrefaction of EFB briquette at 300 °C for 30 minutes. The results showed that mass yield, 78.90 % and energy yield, 64.27 % for EFB Briquette is significantly greater than the mass yield, 24.18 % and energy yield, 56 % observed for EFB in their study. This confirmed that pelletization improves the product yield of torrefaction; increasing the mass yield and energy yield of EFB by a factor of 3 and 1.2 respectively. However, the difference in mass and energy yields of EFB and EFB briquette compared here could also be due to two factors. Primarily, the solid compact nature of EFB briquettes ensures a slower rate pyrolysis or devolatization compared to EFB. Secondly, the difference may be due to the torrefaction technique used; which also presupposes that biomass torrefaction in inert environment may be a limiting factor. The results in Table 1 also reveal that the energy density of EFB briquettes increased from 1.05 to 1.49 during torrefaction primarily due to the increase in the heating value of the torrefied products.

**Appearance of torrefaction products**

Figure 2 presents the raw feedstock oil palm EFB briquettes and torrefied products at different torrefaction temperatures. The samples are denoted; **a** – Raw (untorrefied) EFB briquette; **b** – Torrefied at 250 ºC **(Torr 250); c** – Torrefied at 275 ºC **(Torr 275)**; **d** – Torrefied at 300 ºC **(Torr 300)**.

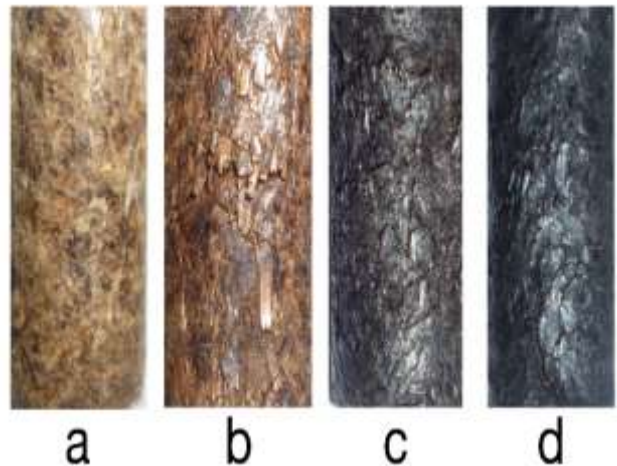

**Fig 2**. Torrefied EFB Briquettes (a) Raw oil palm EFB Briquettes; (b) Torrefied at 250 ºC; (c) Torrefied at 275 ºC; (d) Torrefied at 300 ºC.

As can be observed, the colour of torrefied EFB briquettes changed progressively from greyish brown to black with increasing temperature from 250 °C to 300 °C. The progressive change in colour EFB briquette is primarily due to the feedstock devolatization and increase in carbon content with increasing temperature during torrefaction. Similar trends were also reported by *Uemura et al., 2011* and *Couhert et al., 2009* [7, 17].

## Chemical Fuel Properties

Table 2 presents an overview of the chemical fuel properties of EFB briquettes before and after torrefaction. The carbon content (C) increased significantly by 40 % after torrefaction while oxygen (O) and hydrogen (H) content decreased.

Table 2. Chemical properties of Torrefied oil palm EFB briquettes.

| Sample | C | H | N | S | O | Formulae |
|---|---|---|---|---|---|---|
| EFB Briq | 45.14 | 6.05 | 0.54 | 0.20 | 48.08 | $C_1H_{1.61}O_{0.80}$ |
| Torr 250 | 49.47 | 5.91 | 0.52 | 0.54 | 43.56 | $C_1H_{1.43}O_{0.66}$ |
| Torr 275 | 54.54 | 5.61 | 0.81 | 0.47 | 38.58 | $C_1H_{1.23}O_{0.53}$ |
| Torr 300 | 63.33 | 5.02 | 1.07 | 0.43 | 30.17 | $C_1H_{0.95}O_{0.36}$ |

The O/C ratio decreased from 1.07 to 0.48 while the H/C ratio decreased from 0.13 to 0.08 during torrefaction. This reportedly accounts for the improved properties of torrefied biomass such as heating value, hygroscopic nature and grindability [9, 10]. However an increase in the sulphur content (0.2 - 0.43) and nitrogen content (0.54 - 1.07) was observed after torrefaction. This can potentially result in increased NOx and SOx emissions during thermal applications for power generation. However a possible justification for the increased N content after torrefaction may be nitrogen based side reactions *Eqn 4* although these reactions typically occur at high temperatures and pressures in the presence of metal catalysts supported on $K_2O$, $CaO$, $SiO_2$ and $Al_2O_3$ [18].

$$N_2 + 3H_2 \leftrightarrow 2NH_3 \tag{4}$$

Consequently, the X-ray fluorescence (XRF) analysis of EFB briquette was carried out and the results confirmed the presence of $K_2O$, $CaO$, $SiO_2$, $Al_2O_3$ and $Fe_2O_3$ in the sample. Hence, these compounds may have catalytically promoted the nitrogen reactions which accounts for the increase the N content in the torrefied samples. Similarly, increase in sulphur content in the torrefied samples may be due to the thermo-catalytic Claus desulphurization process. *Dowling et al.,* (1990) showed that the reaction of hydrogen and sulphur can occur under conditions similar to the Claus process, *Eqns 5-6* [19]. Alternatively, the partial oxidation of hydrogen sulphide, *Eqn 7*, can yield hydrogen, water, and elemental sulphur at 400 °C over alumina catalysts [20].

$$H_2 + \left(\frac{1}{2}\right)S_2 \leftrightarrow H_2S \tag{5}$$

$$2H_2S + SO_2 \leftrightarrow \left(\frac{3}{2}\right)S_2 + 2H_2O \tag{6}$$

$$H_2S + \frac{1}{2}O_2 \rightarrow H_2O + \frac{1}{8}S_8 \tag{7}$$

Furthermore, the exothermic partial oxidation reactions in *Eqns 5&7* usually occur at high temperatures, however the presence of catalytic inorganic species $K_2O$, $CaO$, $SiO_2$, $Al_2O_3$ and $Fe_2O_3$ may have influenced the process of sulphur deposition and increased S content at the torrefaction temperatures examined.

## Calorific value of torrefaction products

The calorific (higher heating value, HHV) value of the torrefied samples is presented in Table 3. The heating value of oil palm EFB 17.43 MJ/kg [5] which is lower than the HHV of EFB briquette is 17.57 MJ/kg observed in this study.

Table 3. Heating value of Torrefied oil palm EFB Briquettes.

| Temperature (°C) | HHV (MJ/kg) | Diff in HHV (MJ/kg) | Δ in HHV (%) |
|---|---|---|---|
| EFB Briq | 17.57 | - | - |
| Torr 250 | 19.72 | 2.15 | 12.24 |
| Torr 275 | 21.35 | 3.78 | 21.51 |
| Torr 300 | 26.24 | 8.67 | 49.35 |

The results showed that the HHV of the EFB briquettes increased from 17.57 MJ/kg to 26.24 MJ/kg after torrefaction from 250 °C to 300 °C. This is primarily due to the increase in carbon, fixed carbon content as well as the decrease in oxygen content in Table2. Therefore, we can infer that the torrefaction process significantly improved the heating value of EFB by 49 %. In addition, the HHV of torrefied EFB briquette, 26.24 MJ/kg, at 300 ºC is higher than lignite and bituminous coal (24.45 MJ/kg) reported in literature [9].

## 4. Conclusion

This torrefaction of oil palm EFB briquettes produced a solid uniform fuel with improved thermochemical and physical properties. The thermal upgrading of the fuel properties was enhanced by compact nature of EFB briquettes ensures a slow rate of pyrolysis during torrefaction. With its improved heating value (HHV) of 26.24 MJ/kg after torrefaction at 300 ºC for 1 hour, oil palm EFB briquettes can potential be utilized for future thermal applications. However, $CO_2$, NOx and SOx recovery systems may be required to improve the overall sustainability of the thermochemical conversion process of torrefied EFB briquettes.

## References

1. Lund, H., Renewable energy strategies for sustainable development. Energy, 2007. **32**(6): p. 912-919.
2. Da Silva, C.G., Renewable energies: Choosing the best options. Energy, 2010. **35**(8): p. 3179-3193.
3. Appel, A.M., Electrochemistry: Catalysis at the boundaries. Nature, 2014. **508**(7497): p. 460-461.
4. Bridgeman, T., et al., Torrefaction of reed canary grass, wheat straw and willow to enhance solid fuel qualities and combustion properties. Fuel, 2008. **87**(6): p. 844-856.
5. Uemura, Y., et al., Torrefaction of oil palm EFB in the presence of oxygen. Fuel, 2013. **103**: p. 156-160.
6. Pimchuai, A., A. Dutta, and P. Basu, Torrefaction of agriculture residue to enhance combustible properties†. Energy & Fuels, 2010. **24**(9): p. 4638-4645.
7. Uemura, Y., et al., Torrefaction of oil palm wastes. Fuel, 2011. **90**(8): p. 2585-2591.
8. Grover, P. and S. Mishra, Biomass briquetting: technology and practices. 1996: Food and Agriculture Organization of the United Nations.
9. Basu, P., Biomass gasification, pyrolysis and torrefaction: practical design and theory. 2013: Academic Press.
10. Bergman, P.C., et al., Torrefaction for biomass co-firing in existing coal-fired power stations. Energy Centre of Netherlands, Report No. ECN-C-05-013, 2005.
11. Arias, B., et al., Influence of torrefaction on the grindability and reactivity of woody biomass. Fuel Processing Technology, 2008. **89**(2): p. 169-175.
12. Abdullah, H. and H. Wu, Biochar as a fuel: 1. Properties and grindability of biochars produced from the pyrolysis of mallee wood under slow-heating conditions. Energy & Fuels, 2009. **23**(8): p. 4174-4181.
13. Uslu, A., A.P. Faaij, and P.C. Bergman, Pre-treatment technologies, and their effect on international bioenergy supply chain logistics. Techno-economic evaluation of torrefaction, fast pyrolysis and pelletisation. Energy, 2008. **33**(8): p. 1206-1223.
14. Repellin, V., et al., Modelling anhydrous weight loss of wood chips during torrefaction in a pilot kiln. Biomass and bioenergy, 2010. **34**(5): p. 602-609.
15. Almeida, G., J. Brito, and P. Perré, Alterations in energy properties of eucalyptus wood and bark subjected to torrefaction: the potential of mass loss as a synthetic indicator. Bioresource technology, 2010. **101**(24): p. 9778-9784.
16. Chen, W.-H. and P.-C. Kuo, A study on torrefaction of various biomass materials and its impact on lignocellulosic structure simulated by a thermogravimetry. Energy, 2010. **35**(6): p. 2580-2586.
17. Couhert, C., S. Salvador, and J. Commandre, Impact of torrefaction on syngas production from wood. Fuel, 2009. **88**(11): p. 2286-2290.
18. Myers, R., The basics of chemistry. 2003: Greenwood Publishing Group.
19. Dowling, N.I., J.B. Hyne, and D.M. Brown, Kinetics of the reaction between hydrogen and sulfur under high-temperature Claus furnace conditions. Industrial & Engineering Chemistry Research, 1990. **29**(12): p. 2327-2332.
20. Clark, P., N. Dowling, and M. Huang, Production of H 2 from catalytic partial oxidation of H 2 S in a short-contact-time reactor. Catalysis Communications, 2004. **5**(12): p. 743-747.